# Enhancement of giant magnetoresistance effect in the Ruddlesden-Popper phase $Sr_3Fe_{2-x}Co_xO_{7-\delta}$: Predominant role of oxygen nonstoichiometry and magnetic phase separation

(Running head: Enhancement of GMR effect in the RP phase $Sr_3Fe_{2-x}Co_xO_{7-\delta}$)

T. Motohashi[1], B. Raveau, M. Hervieu, A. Maignan, V. Pralong,

N. Nguyen, and V. Caignaert

*Laboratoire CRISMAT, UMR CNRS ENSICAEN 6508, 6 bd Maréchal Juin*

*14050 CAEN Cedex 4 France*

(Dated: Nov. 1, 2005, revised manuscript: Dec. 23, 2005)

The magnetic and magnetotransport properties of the $Sr_3Fe_{2-x}Co_xO_{7-\delta}$ system ($0.2 \leq x \leq 1.0$) were systematically investigated. This oxide system exhibits a giant magnetoresistance (GMR) effect at low temperatures, reaching up to 80% in 7 T at 5 K. Ac-susceptibility measurements show that there exists a strong competition between ferromagnetic (F) and spin glass states, and the balance between these two magnetic states can be controlled by varying cobalt ($x$) and/or oxygen contents ($\delta$). Importantly, the MR effect is closely related to the magnetic property: the development of magnetic disordering leads to enhancement in the negative MR effect. It is suggested that the compound segregates into F clusters embedded in a non-F matrix, being a naturally occurring analog of the artificial granular-GMR materials, as in the doped perovskite cobaltites, $La_{1-x}Sr_xCoO_3$ ($x < 0.18$).

[1]Corresponding author
Present address: Materials and Structures Laboratory, Tokyo Institute of Technology, 4259 Nagatsuta, Midori-ku, Yokohama 226-8503, Japan
E-mail: t-mot@msl.titech.ac.jp, Phone: +81-45-924-5318, Fax: +81-45-924-5339

**1. Introduction**

Recent investigations on strongly correlated oxides such as underdoped high-$T_c$ superconducting cuprates [1] and colossal magnetoresistance (CMR) manganites [2-3] have revealed that these materials exhibit spatial coexistence of two distinct electronic phases, i.e. a phase separated state within a fine length scale of several tens of nanometers. In such systems, the phase separation is at the origin of the competition between two different states, i.e. superconducting / pseudogapped phases in high-$T_c$ cuprates and ferromagnetic / charge-ordered phases in CMR manganites. Thus, these experimental observations demonstrate that the phase competition plays a crucial role in the novel properties of the strongly correlated electron systems.

Among oxide systems with strong electronic/magnetic correlations, only few studies have been devoted to the cobaltite systems. In this respect, a study on the perovskite $La_{1-x}Sr_xCoO_3$ by Wu and Leighton [4] is of great interest since it shows for the first time the existence of magnetic phase separation in a cobaltite. Recently, the perovskites $SrCo_{1-x}M_xO_{3-\delta}$ ($M$ = Nb, Ru) were also found to involve a strong competition between two magnetic states [5-6]. All these cobaltites exhibit a negative giant magnetoresistance (GMR) at low temperatures, concomitantly with a spin glass (SG) behavior. In fact, small angle neutron scattering experiments on $La_{1-x}Sr_xCoO_3$ by Wu *et al.* [7] show a granular heterostructure which is the likely source of the enhanced negative GMR, in contrast to the CMR effect in manganites, which originates from the double-exchange interaction mechanism. This cobaltite is believed to segregate into metallic ferromagnetic (F) clusters embedded in a non-F insulating matrix, corresponding to a "naturally-created" intergranular GMR material.

These results suggest the possibility of tailoring new GMR materials which possess an "intrinsic" heterostructure originated from a phase competition. For this issue, the layered iron-cobalt oxide, $Sr_3Fe_{2-x}Co_xO_{7-\delta}$, may be one of the most promising candidates. This compound is a member of the Ruddlesden-Popper (RP) oxides with the general chemical formula, $Sr_{n+1}(Fe,Co)_nO_{3n+1}$ ($n = 2$). Previous studies on the magnetic and magnetotransport properties of $Sr_3Fe_{2-x}Co_xO_{7-\delta}$ ($0 \leq x \leq 1$) indicated [8-11] that the title compound shows a disordered magnetic behavior and a large negative MR effect in a wide temperature range below $T_C$. From the experimental fact that the metallic F state is drastically suppressed by decreasing cobalt content ($x$) [10, 12], metallicity and ferromagnetism are suggested to originate mainly from tetravalent cobalt species. Nevertheless the value of MR remains much smaller than that observed for $La_{1-x}Sr_xCoO_3$, reaching a maximum of 47% at 5 K [8-11].

Bearing in mind that the effect of oxygen nonstoichiometry on the magnetic and magnetotransport properties of $Sr_3Fe_{2-x}Co_xO_{7-\delta}$ remains unclear, we have revisited this system. We report herein the detailed magnetic and magnetotransport properties of this system for $0.2 \leq x \leq 1.0$. By optimizing the chemical composition and oxygen content, the value of MR can be enhanced up to 80% in 7 T at 5 K. Our ac-susceptibility data clearly demonstrate that there exists a strong competition between F and SG states, involving most likely phase separation into F clusters embedded in a non-F matrix. We also show that the MR effect is closely related to the magnetic property: the development of magnetic disordering leads to enhancement in the negative MR effect, being consistent with the heterostructure model as in the perovskite cobaltites, $La_{1-x}Sr_xCoO_3$ ($x < 0.18$).

## 2. Experimental

A sample series of $Sr_3Fe_{2-x}Co_xO_{7-\delta}$ ($0.2 \leq x \leq 1.0$) was prepared by a conventional solid-state reaction. Powder mixtures of $SrCO_3$, $Fe_2O_3$, and $Co_3O_4$ with appropriate ratios were calcined in flowing $O_2$ gas at 800°C for 12 h. The calcined products were ground, pelletized, and fired in flowing $O_2$ gas at 1200°C for 12 h, followed by slow cooling to room temperature. The present compound was found to experience chemical instability by prolonged exposure to air due to high reactivity with atmospheric water [10-12]. We thus paid the closest attention to prevent sample degradation during the experimental procedures. The sintered pellets were put in a vacuum chamber immediately after taking them out of an atmospheric-controlled furnace and then stored in a glovebox in which water concentration was kept less than 0.05 ppm. By employing this procedure, the samples could be kept for several weeks without any alteration of their properties. A part of the samples was subsequently post-annealed in an oxygen pressure of 10 MPa at 500°C for 24 h.

X-ray powder diffraction (XRPD) analysis was performed using a Philips X-pert Pro diffractometer (Cu $K_\alpha$ radiation). The samples were pulverized in the glovebox and then immediately mounted to the diffractometer such that sample degradation could be minimized. The data were collected in an angular range of $4° \leq 2\theta \leq 120°$ with a relatively fast scan rate of 5° / min. The electron diffraction (ED) and energy dispersive spectroscopy (EDS) analyses were performed using a transmission electron microscope (TEM), JEOL 200CX equipped with a Kevex EDS analyzer. The crystallites of the as-synthesized $x = 1.0$ sample were gently crushed in the glovebox using $CCl_4$ as a suspension liquid. A certain amount of flakes was deposited onto a holey carbon film, supported by copper grid. The grid was handled under a dry atmosphere and put on the sample holder in the glovebox. The EDS analyses were carried out on several tens crystallites. The oxygen content, 7-$\delta$, was determined by the cerimetric titration technique. Note that the iodometric titration, which has been generally applied for various oxide materials, does not give accurate results due to a small difference in the equilibrium potential between $Fe^{3+}/Fe^{2+}$ ($E^0$ = 0.77 V) and $2I^-/I_2$ (0.54 V) redox couples [13]. The oxidation states of iron were determined from Mössbauer spectroscopy. Mössbauer resonance spectra were obtained with transmission geometry at room temperature using a $\gamma$-ray source from $^{57}$Co embedded in the rhodium matrix. The isomer shift was referred to metallic $\alpha$-Fe.

Magnetic and magnetotransport properties were investigated using a PPMS facility (Quantum Design). The dc-magnetization ($M$) measurements were carried out with the dc extraction method in a temperature range of 5 K $\leq T \leq$ 300 K and in magnetic fields up to 5 T. The ac-susceptibility ($\chi$) was measured in an ac-field of 3 Oe with frequencies of 10, $10^2$, $10^3$, and $10^4$ Hz. Resistivity ($\rho$) and magnetoresistance (MR) measurements were performed with a four-point method in a magnetic field of 0 – 7 T. The sample pellets were installed in a helium-filled cryostat immediately after preparing indium electrodes at the sample surface (within 30 minutes). The MR vs $H$ curves were taken at several temperatures. In each measurement, the samples were first zero-field cooled and then the applied field was increased from 0 to 7 T and decreased back to 0 T, and further to -7 T.

## 3. Results

Single-phase samples of $Sr_3Fe_{2-x}Co_xO_{7-\delta}$ were successfully obtained for a wide range of chemical composition, 0.2 $\leq x \leq$ 1.0. XRPD analyses indicated that all the diffraction peaks of the samples are indexed based on the tetragonal space group *I*4/*mmm* as exemplified for $x = 0.5$ (as-synthesized) in Fig. 1(a). Both of the $a$- and $c$-axis lengths first decrease with increasing the cobalt content ($x$), and then exhibit a saturating behavior for larger $x$ ($\geq 0.5$) as shown in Fig. 2. Also, these values are decreased by oxygen pressure annealing for instance from $a = 3.854(0)$ Å and $c = 20.105(0)$ Å to $a = 3.851(0)$ Å and $c = 20.088(0)$ Å for $x = 0.5$. These results are consistent with those previously observed for the present compound [10, 12].

The ED study confirmed good crystallinity of the sample. The EDS analyses on the as-synthesized $x = 1.0$ sample revealed the homogeneity of the sample, with a cationic ratio of Sr/Fe/Co = 3/1/1, in good agreement with the nominal one. The reconstruction of the reciprocal space, carried out by rotating along the crystallographic axes in the ED analyses, showed that a set of intense Bragg peaks is consistent with the $I4/mmm$ tetragonal structure. However, weak additional spots are observed as illustrated in the [001] and [010] ED patterns [Figs. 1(b) and 1(c), respectively]. These extra spots violate the diffraction condition ($hk0$, $h+k = 2n$) of the $I4/mmm$ space group [e.g. a 010 reflection marked with a curved arrow in Fig. 1(b)], involving a lowering of the symmetry toward an $A$-type orthorhombic space group. Another set of weak extra reflections are also observed on incommensurate positions [see white arrows in Figs. 1(b) and 1(c)]. They evidence a modulated structure with a modulation vector $\vec{q}^* = \alpha \vec{a}^*$ with $\alpha \approx 0.52$. The ED patterns [Figs. 1(b), 1(c), and 1(d)] are indexed using four $hklm$ indices. Note that similar spots have been observed along the perpendicular ($\vec{b}^*$) direction, as a consequence of the formation of twinning domains. The latter set of the extra reflections forms diffuse streaks along $\vec{c}^*$ in the [010] ED pattern [Fig. 1(c)]. One may notice that the diffuse streaks exhibit an undulation component along $\vec{a}^*$, consistent with the incommensurate character of the modulated structure with $\alpha \approx 0.52$. The indexation of this ED pattern is schematically given in Fig. 1(d). These observations suggest the existence of ordering phenomena along the equivalent $\{100\}_{sub}$ direction of the tetragonal subcell, likely associated with oxygen vacancy ordering within the double perovskite layers. The diffuse streaks along $\vec{c}^*$ indicate that the superstructure is not perfectly established in the present sample but rather of short range. Despite the high degree of disorder, the observations suggest $Bmmm(\alpha 00)000$ as a possible super space group. It is noteworthy pointing out that the direction of the superstructure, namely $\vec{a}_p$, is consistent with the formation of tetragonal pyramids $(Fe,Co)O_5$ rather than tetrahedra as in the brownmillerite-type derivatives. From this hypothesis, it is straightforward that the $O_2$-annealing process is crucial in the stabilization of long range ordering. It is of interest to clarify the effect of oxygen nonstoichiometry on the crystal structure, but this is not the aim of the present paper.

Oxygen content values, "7-δ", determined by cerimetric titration are summarized in Table 1. One observes a certain level of oxygen vacancies in all the samples but the δ value does not depend strongly on the Fe/Co ratio. It should be noted that the δ value decreases and thereby the Fe/Co valency increases after oxygen annealing. To obtain information on the iron valency, powder $^{57}$Fe Mössbauer spectroscopy was performed on the as-synthesized and $O_2$-annealed samples with $x = 0.5$ and 0.8. All of the samples studied show a paramagnetic behavior at room temperature (Fig. 3). The hyperfine parameters determined by least-square fits are given in Table 2. The isomer shift (*IS*) values allow to characterize two different iron components in spite of the existence of only one crystallographic Fe/Co site: the first component is clearly attributed to $Fe^{4+}$ owing to its negative *IS* value ($-0.08 \leq IS \leq -0.04$ mm/s), while the *IS* value of the second component ($0.17 \leq IS \leq 0.24$ mm/s) is rather small with respect to that usually observed for $Fe^{3+}$ in octahedral sites (> 0.35 mm/s) [14]. Thus, the oxidation state, 3+$y$, of the latter component cannot be exactly determined. Nevertheless, taking into account the typical *IS* values for $Fe^{4+}$ and $Fe^{3+}$ in octahedral coordination ($\approx$ −0.10 and $\approx$ 0.35 mm/s, respectively), we could suggest that the *y* value is in the range of 0.25 – 0.40. For these compounds, the $Fe^{4+}$ content is much higher in the $O_2$-annealed samples than the as-synthesized samples. The Mössbauer and cerimetric titration results indicate that a large number of high valent cations, i.e. $Fe^{4+}$ or $Co^{4+}$, exists in the lattice.

All the samples show a semiconductive behavior (i.e. $d\rho / dT < 0$) in the whole temperature range (representative data are presented in Fig. 4). Nevertheless, the resistivity neither follows the Arrhenius law nor diverges at lower temperatures, suggesting that the semiconductive upturn does not simply originate from a real-gap opening. The absolute value of $\rho$ systematically and dramatically decreases with increasing Co content (*x*), while the $\rho$ value is reduced by oxygen annealing. This result implies that high valent cobalt ($Co^{4+}$) induces a metallic conductivity as in the perovskite cobaltite $La_{1-x}Sr_xCoO_3$ [7, 15].

It can be seen that the 7 T curve starts to deviate from the 0 T curve around 200 K. Thus, a large negative MR effect definitely appears in a low temperature region. Shown in Fig. 5 is a typical result of MR vs *H* measurements at various temperatures. From this figure, several significant features can be pointed out. First, a negative GMR effect is observed in a wide temperature range. Second, the magnitude of MR increases with lowering temperature, in particular the MR effect is remarkable at temperatures below 50 K (see also the inset of Fig. 4). Third, a large hysteresis is seen only at 5 and 10 K. The 5 K and 10 K curves show resistivity maxima at $H \approx$ −0.7 and −0.3 T, respectively, which correspond to the coercive field ($H_c$) extracted from the *M* vs *H* loops.

To elucidate the effect of the cobalt- or oxygen-content variation on the MR property of $Sr_3Fe_{2-x}Co_xO_{7-\delta}$, magnetotransport measurements were systematically carried out. In Fig. 6(a), the magnitude of MR (in 7 T at 5 K) is plotted as a function of cobalt content ($x$) for the as-synthesized and $O_2$-annealed sample series (marked with open and closed circles, respectively). Apparently, the MR property correlates with the chemical composition: i.e. the MR value systematically increases with decreasing $x$, reaching 66% and 80% in 5 T and 7 T, respectively, for the as-synthesized $x = 0.5$ sample. These values are record-high and much larger than that ever reported for the present compound, e.g. 47% in 5 T for $x = 0.5$ [10] or 47% in 7 T for $x = 0.6$ [10]. On the other hand, the MR value is significantly lowered by oxygen annealing in the whole cobalt compositions. In this figure, results of highly oxygenated samples by Ghosh and Adler [10], annealed in $P(O_2)$ = 400 – 850 bar, are also presented (closed triangles). It can be seen that their samples follow the same trend: i.e. the increase in the oxygen content (or valence number of Fe/Co) results in deterioration of the MR characteristics. In some cobalt oxides, the negative MR effect was found to be enhanced when the localized character of the material becomes more pronounced [16]. For $Sr_3Fe_{2-x}Co_xO_{7-\delta}$, the absolute value of $\rho$ at 5 K dramatically increases with decreasing cobalt content ($x$), and highly semiconductive samples (with smaller $x$) indeed tend to show larger MR, as seen in Figs. 6(a) and 6(b). From this point of view, one might simply consider that the reduced MR effect in the oxygenated samples is due to smaller degree of carrier localization. To verify this statement, we plotted the MR value in 7 T at 5 K against the $\rho$ value at 5 K [inset of Fig. 6(b)]. From this figure, it is deduced that the MR magnitude increases as the $\rho_{5K}$ value increases. Nevertheless, two regimes can be distinguished with a greater dependence of MR on $\rho_{5K}$, as $\rho_{5K}$ becomes smaller than 10 $\Omega$ cm.

Due to the strong interplay between charges and spins in transition-metal oxides, it is thus highly necessary to see how the cobalt content and oxygen nonstoichiometry influence the magnetic behavior of $Sr_3Fe_{2-x}Co_xO_{7-\delta}$. The $M$ vs $T$ curves, measured in a magnetic field of 0.3 T, show a significant increase in $M$ as the temperature decreases (Fig. 7), indicating the presence of F interactions. The temperature at which magnetization starts to increase does not depend strongly on the Co content ($x$), i.e. $T \approx 200$ K. Note that this temperature coincides with the onset temperature of the negative GMR effect. The magnitude of $M$ systematically increases as both the cobalt and oxygen contents increase. These results, suggesting that ferromagnetism originates mainly from the presence of tetravalent cobalt, are in good agreement with the previous observations [10-12]. On the other hand, the $M$ vs $H$ loops (Fig. 7) show a steep increase in $M$ at low fields and a large magnetization reaching ~5 $\mu_B$ / f.u. in 5 T, as typically seen in F materials. Nevertheless, the magnetization does not completely saturate even in 5 T and the non-saturating behavior is more pronounced for Co-poor and/or less-oxygenated samples. This clearly indicates that the samples are magnetically inhomogeneous, most likely involving magnetic disordering such as a SG state.

It is well known that ac-susceptibility measurements provide valuable information on the magnetic disordering phenomena. Shown in Fig. 8 are the in-phase ($\chi'$; upper panel) and out-of-phase components ($\chi''$; lower panel) of ac-susceptibility of the $Sr_3Fe_{2-x}Co_xO_{7-\delta}$ sample series. For the as-synthesized $x = 0.2$ and $0.5$ samples, $\chi'$ and $\chi''$ are small in magnitude and show a sharp cusp around 40-50 K, which is strongly frequency dependent and shifts toward higher temperatures as the frequency ($f$) increases. This is a typical feature of SG materials [17]. As the cobalt content increases, i.e. $x = 0.8$ and 1.0, the magnitude of $\chi'$ rapidly increases and the cusp becomes significantly broadened. Moreover, in the χ"(T) plot an additional peak appears around 80 K and its magnitude systematically increases with increasing Co content. Importantly, this high-temperature peak is $f$ independent exhibiting a sharp contrast to the low-temperature peak. The high-temperature component is more pronounced in the $O_2$-annealed samples than in the as-synthesized samples in the whole cobalt compositions. This result clearly demonstrates that two different magnetic states coexist and compete with each other. Most likely, this strong magnetic competition, involving phase separation, plays an important role in the large negative MR effect.

## 4. Discussion

It is natural to ascribe the $f$-independent χ" peak (at ≈80 K) to the onset of ferromagnetism, and the $f$-dependent peak (at 40-50 K) to a SG freezing temperature, that is, the magnetic behavior observed here is of re-entrant SG like. To obtain deeper understanding of the latter ($f$-dependent) component, we applied the standard theory for dynamical scaling near a phase transition [17-18] as described by

$$\frac{\tau}{\tau_0} = \left(\frac{T_f - T_{SG}}{T_{SG}}\right)^{-z\nu} \quad (1)$$

where $\tau = f^{-1}$, $T_f$ is the $f$-dependent freezing temperature determined by the inflection point in χ"(T) or maximum in χ'(T), $T_{SG}$ is the critical temperature for SG formation (this is equivalent to the $f \to 0$ value of $T_f$), $\tau_0$ is the characteristic time scale for the spin dynamics, and $z\nu$ is a constant term called "dynamic exponent". As represented by a result in Fig. 9, the frequency dependence of $T_f$ is well fitted to Eq. (1) for all of the samples. The $z\nu$ value, ranging from 4.2 to 5.4, should be compared to that reported for canonical SG systems, i.e. $z\nu = 5.5$ for a CuMn alloy [17]. In Fig. 10, $\tau_0$ and $T_{SG}$ obtained by least-square calculations are plotted as a function of cobalt content ($x$). The magnitude of $\tau_0$ of the present compound, $10^{-9} \sim 10^{-7}$ s, is much larger than that reported for canonical spin-glass systems, $\tau_0 \sim 10^{-12}$ s [17]. Such a large $\tau_0$ value implies the existence of strong F interactions even in the glassy state. It should be noted that $\tau_0$ tends to increase with increasing cobalt content and $O_2$ annealing, consistent with the experimental fact that ferromagnetism becomes predominant in such Co-rich and/or $O_2$-annealed samples [19].

It is noteworthy pointing out that $Sr_3Fe_{2-x}Co_xO_{7-\delta}$ exhibits remarkable similarities to the "intergranular-GMR" perovskite cobaltites, $La_{1-x}Sr_xCoO_3$ ($x < 0.18$) [4, 7, 15]. Both compounds indeed show (i) a disordered magnetic behavior leading to a SG or a cluster glass state, (ii) a large negative MR effect in the whole temperature range below $T_C$, and (iii) a distinct irreversibility in the MR vs $H$ curves which is closely related to the $M(H)$ loop. For the perovskite cobaltites, small-angle neutron scattering experiments have shown [7] that the material segregates into F metallic clusters embedded in a non-F insulating matrix, being a naturally occurring analog of the artificial granular-GMR materials [20]. We suggest that the magnetic phase separation in $Sr_3Fe_{2-x}Co_xO_{7-\delta}$ creates a similar heterostructure. In fact, the magnetic and MR properties of the present compound can be consistently explained upon assuming the heterostructure. The re-entrant SG behavior could be a consequence of formation of F clusters at $\approx$ 80 K and subsequent freezing of spin alignment of the clusters at 40-50 K. Application of a magnetic field at low temperatures produces ferromagnetic alignment of the clusters. The negative MR effect arises from reduction of spin-dependent scattering of carriers between the clusters in applied magnetic fields. We note that such a heterostructure picture has also been proposed for $Sr_3Fe_{2-x}Co_xO_{7-\delta}$ by Sánchez-Andújar *et al.* [11]. The authors of the paper investigated the property of an $Sr_3Fe_{1.5}Co_{0.5}O_{6.67}$ sample and suggested that the delocalizing character of cobalt brings about strong F interactions, resulting in the formation of Co-containing F clusters, embedded in the Fe-rich antiferromagnetic (AF) matrix.

The ac-susceptibility measurements (Fig. 8) evidence a strong competition between F and SG states in the present system. This result also shows that the balance between these two magnetic states (in other words, the degree of magnetic disordering) can be easily controlled by varying cobalt and/or oxygen contents. Importantly, we see that the MR effect is closely related to the magnetic property: i.e. the F behavior becomes predominant with increasing cobalt and/or oxygen contents, and the enhanced F interaction leads to deterioration of the negative MR effect. Within the granular heterostructure model [7], the dominance of ferromagnetism in Co-rich and/or $O_2$-annealed samples may reflect the increased size or volume fraction of F clusters, as deduced from the increase in $\tau_0$ (Fig. 10). In such a situation, we expect that the F region is well percolated over the sample and the probability of two F clusters to isolate each other gets lowered, resulting in a *large* conductivity in zero field and a small change in resistivity by applying fields, i.e. a *small* negative MR effect. These statements are in perfect agreement with those observed in the present system. Small angle neutron scattering experiments will allow to establish the veracity of our model. We recognize the importance of studies on Co-rich ($x > 1.0$) and oxygen-depleted samples to further control the MR characteristic of the present system. However, our efforts to prepare samples for magnetotransport measurements have been unsuccessful so far due to the poor chemical stability against ambient atmosphere of Co-rich samples.

Let us finally discuss the MR property of $Sr_3Fe_{2-x}Co_xO_{7-\delta}$ in comparison with the related system, a simple (3D) perovskite $SrFe_{1-y}Co_yO_{3-\delta}$. First, within the 3D oxide, the MR mechanisms are already very complex. For instance the magnetic structures of the end members corresponding to $y = 0.0$ and $1.0$ are helical and ferromagnetic, respectively. Their MR behavior differs greatly, since no MR has been reported yet for $SrCoO_3$, whereas in $SrFeO_{3-\delta}$ the sign and magnitude of MR has been shown to depend on the oxygen content, the best result being 90% in 9 T at 70 K for $SrFeO_{2.85}$ [21]. Interestingly, substitution of cobalt for iron in $SrFeO_{3-\delta}$ induces ferromagnetism with the maximum values of $T_C$ and $M$ at $y = 0.5$ [22-23]. This is in agreement with that observed for the $Sr_2Fe_{2-x}Co_xO_{7-\delta}$ series for which the F behavior is enhanced as $x$ increases. In that respect a strong covalency between $Fe^{4+}(Co^{4+})$-O-$Fe^{4+}(Co^{4+})$ is responsible for both ferromagnetism and metallicity. Accordingly, decreasing the population of tetravalent species by decreasing oxygen content strongly affects these properties by creating antiferromagnetic interactions which lead to magnetic phase separation. However, in this RP $n = 2$-member, $Sr_3Fe_{2-x}Co_xO_{7-\delta}$, it must be pointed out that the 2D character of the present compound may also play a role in the enhanced MR effect as for $(La,Sr)_3Mn_2O_7$ with a layered RP structure [24]. It was shown that for the RP manganites the carrier tunneling along the layer stacking direction, i.e. through natural $[(La,Sr)O]_2$ rock-salt-type barriers, is favored upon magnetic field application, leading to an additional MR effect, besides the conventional CMR effect of manganite perovskites. Again the presence of oxygen vacancies may also affect the transverse resistivity and thus change this intrinsic tunneling mechanism. At that point, it is clear that neutron diffraction studies should be performed to locate the oxygen vacancies in the structure. Also, in-plane and out-of-plane transport measurements on single crystalline samples are strongly needed to conclude the origin of the enhanced MR and to discard the grain boundary effect.

## 5. Conclusions

In the present study, the magnetic and magnetotransport properties of the $Sr_3Fe_{2-x}Co_xO_{7-\delta}$ system ($0.2 \leq x \leq 1.0$) were systematically investigated. It is demonstrated that this oxide system exhibits a giant magnetoresistance (GMR) effect at low temperatures, reaching up to 80% in 7 T at 5 K. The MR property is characterized by (i) a large MR effect in a wide temperature range, (ii) a systematic increase in MR with lowering temperature, and (iii) a large hysteresis at lowest temperatures which is strongly related to the $M$ vs $H$ loop. The ac-susceptibility data evidence that there exists a strong competition between ferromagnetic (F) and spin glass (SG) states, and the balance between the two magnetic states can be controlled by varying cobalt ($x$) and/or oxygen contents ($\delta$). Importantly, the MR effect is closely related to the magnetic property: the F behavior becomes predominant with increasing cobalt and/or oxygen contents, and the enhanced F interaction leads to deterioration of the negative MR effect. We suggest that the compound segregates into F clusters embedded in a non-F matrix, being a naturally occurring analog of the artificial granular-GMR materials, as in the doped perovskite cobaltites, $La_{1-x}Sr_xCoO_3$ ($x < 0.18$). Additionally, intrinsic interlayer tunneling may also play a role in the enhanced MR effect.

Table 1. Oxygen content 7-δ and average valence number of [Co,Fe] ($V_{Co,Fe}$) of the $Sr_3Fe_{2-x}Co_xO_{7-\delta}$ samples, determined by cerimetric titration.

| Sample $Sr_3Fe_{2-x}Co_xO_{7-\delta}$ | Oxygen content 7-δ | Valence number $V_{Co,Fe}$ |
|---|---|---|
| $x = 0.3$ as-synthesized | 6.72 | 3.72 |
| $x = 0.5$ as-synthesized | 6.74 | 3.74 |
| $x = 0.8$ as-synthesized | 6.67 | 3.67 |
| $x = 1.0$ as-synthesized | 6.70 | 3.70 |
| $x = 0.3$ $O_2$-annealed | 6.78 | 3.78 |
| $x = 0.5$ $O_2$-annealed | 6.83 | 3.83 |
| $x = 0.8$ $O_2$-annealed | 6.69 | 3.69 |
| $x = 1.0$ $O_2$-annealed | 6.71 | 3.71 |

Table 2. Hyperfine Mössbauer parameters of the as-synthesized and $O_2$-annealed $Sr_3Fe_{2-x}Co_xO_{7-\delta}$ samples with $x = 0.5$ and 0.8 at room temperature. The statistical errors of the isomer shift (*IS*), quadrupole splitting (*QS*), and relative intensity *I*, are ±0.01 mm/s, ±0.02 mm/s, and ±2%, respectively.

| Sample | Isomer shift[a] *IS* (mm/s) | Quadrupole splitting *QS* (mm/s) | Relative intensity *I* (%) | Iron valency |
|---|---|---|---|---|
| $x = 0.5$ as-synthesized | −0.08 | 0.10 | 41 | +4 |
| | 0.20 | 0.25 | 59 | +3+*y* |
| $x = 0.5$ $O_2$-annealed | −0.04 | 0.11 | 64 | +4 |
| | 0.23 | 0.14 | 36 | +3+*y* |
| $x = 0.8$ as-synthesized | −0.08 | 0.08 | 37 | +4 |
| | 0.18 | 0.25 | 63 | +3+*y* |
| $x = 0.8$ $O_2$-annealed | −0.05 | 0.12 | 57 | +4 |
| | 0.21 | 0.16 | 43 | +3+*y* |

[a] The value relative to metallic α-Fe.

**Figure Captions**

Fig. 1
(a) X-ray powder diffraction pattern for the as-synthesized $Sr_3Fe_{1.5}Co_{0.5}O_{7-\delta}$ sample. The pattern is indexed in the *I*4/*mmm* space group (No. 139) with $a = 3.854$ Å, $c = 20.11$ Å.
(b) [001] ED pattern of the same sample. A 010 extra reflection is indicated with a curved arrow. Weak reflections due to an incommensurate modulated structure are marked with white arrows. (c) [010] ED pattern of the same sample. Diffuse streaks are marked with white arrows. (d) A schematic drawing of the experimental ED pattern, explaining the superimposition of the [100] and [010] patterns due to twinning phenomena. It shows the origin of the two streaky diffuse lines.

Fig. 2
Lattice parameters *a* and *c* of the tetragonal unit cells refined from X-ray powder diffraction patterns of the $Sr_3Fe_{2-x}Co_xO_{7-\delta}$ samples. Open (blue) and solid (red) symbols denote the data of the as-synthesized and $O_2$-annealed samples, respectively.

Fig. 3
$^{57}$Fe Mössbauer spectra of the as-synthesized and $O_2$-annelaed $Sr_3Fe_{2-x}Co_xO_{7-\delta}$ samples (*x* = 0.5 and 0.8) in a paramagnetic state at room temperature.

Fig. 4
Temperature dependence of resistivity ($\rho$) for the as-synthesized $Sr_3Fe_{2-x}Co_xO_{7-\delta}$ samples with *x* = 0.5 (red curves) and 0.8 (blue curves). Solid and broken curves show $\rho$ data recorded in 0 T and 7 T, respectively. The inset shows the MR value (in 7 T) of these samples as a function of temperature. The MR values were obtained from the MR vs *H* measurements.

Fig. 5
MR vs *H* plots for the as-synthesized *x* = 0.5 sample at 5, 10, 50, 100, 150, and 200 K. In each measurement, the sample was zero-field-cooled, and the magnetic field was swept from 0 to 7 T, and then from 7 to -7 T. A large hysteresis is seen only at 5 and 10 K.

Fig. 6
(a) The MR value (at 5 K, 7 T) of the $Sr_3Fe_{2-x}Co_xO_{7-\delta}$ samples as a function of cobalt

content ($x$). Open (blue) and solid (red) circles are data plots of the as-synthesized and $O_2$-annealed samples, respectively. MR data of highly-oxygenated and less-oxygenated samples by Ghosh and Adler [10] are plotted with solid and open (black) triangles, respectively. Results of Bréard *et al.* (solid diamond) [9] and Sánchez-Andújar *et al.* (cross) [11] are also given in this figure. Solid and broken lines are guides to eye. (b) The $\rho$ value (at 5 K, 0 T) of the $Sr_3Fe_{2-x}Co_xO_{7-\delta}$ samples as a function of cobalt content ($x$). The inset shows the MR value (at 5 K, 7 T) against the $\rho$ value (at 5 K, 0 T) of our $Sr_3Fe_{2-x}Co_xO_{7-\delta}$ samples. Solid and broken lines are guides to eye.

Fig. 7

(Upper panels) Temperature dependence of dc-magnetization ($M$) of the $Sr_3Fe_{2-x}Co_xO_{7-\delta}$ samples with $x = 0.5, 0.8$, and $1.0$. The data were recorded in an applying field of 0.3 T in the zero-field-cooling (solid symbols) and field-cooling (open symbols) modes. Red triangles and blue circles represent the data plots for the as-synthesized and $O_2$-annealed samples, respectively. (Lower panels) $M$ vs $H$ loops at 5 K for the as-synthesized (blue triangles) and $O_2$-annealed (red circles) samples.

Fig. 8

Temperature dependence of ac-susceptibility, $\chi'$ (upper panels) and $\chi''$ (lower panels) for the $Sr_3Fe_{2-x}Co_xO_{7-\delta}$ samples with $x = 0.2, 0.5, 0.8$, and $1.0$. The data were recorded in an ac-field of 3 Oe with frequencies of 10, $10^2$, $10^3$, and $10^4$ Hz. Open and solid symbols denote the data plots for the as-synthesized and $O_2$-annealed samples, respectively.

Fig. 9

$\log f$ vs $\log[(T_f - T_{SG}) / T_{SG}]$ plots for the as-synthesized $x = 0.5$ sample. A solid line is the best fit to the data with the parameters shown in the figure.

Fig. 10

Dependence of the characteristic time for the spin dynamics ($\tau_0$) and the spin-glass freezing temperature ($T_{SG}$) on cobalt content ($x$) in $Sr_3Fe_{2-x}Co_xO_{7-\delta}$. These values were obtained by least-square calculations, as represented in Fig. 8. Open (blue) and solid (red) symbols denote the data of the as-synthesized and $O_2$-annealed samples, respectively.

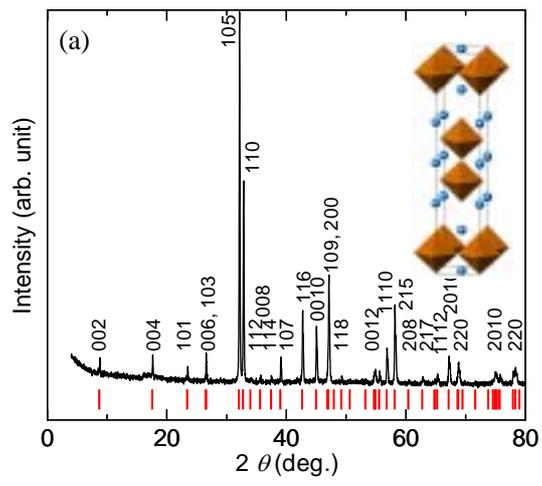

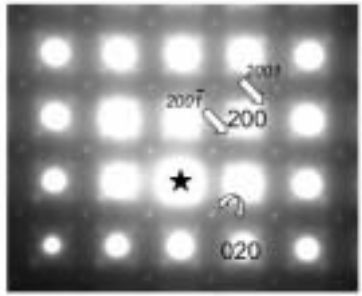

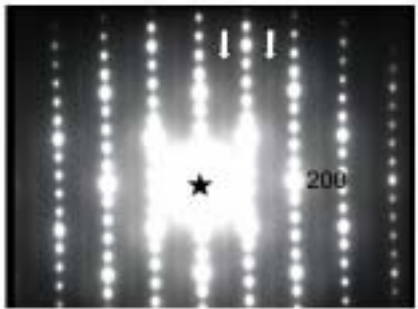

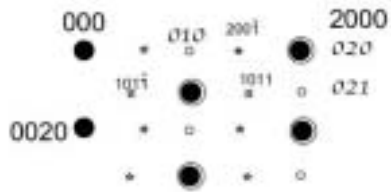

Fig. 1. Motohashi *et al.*

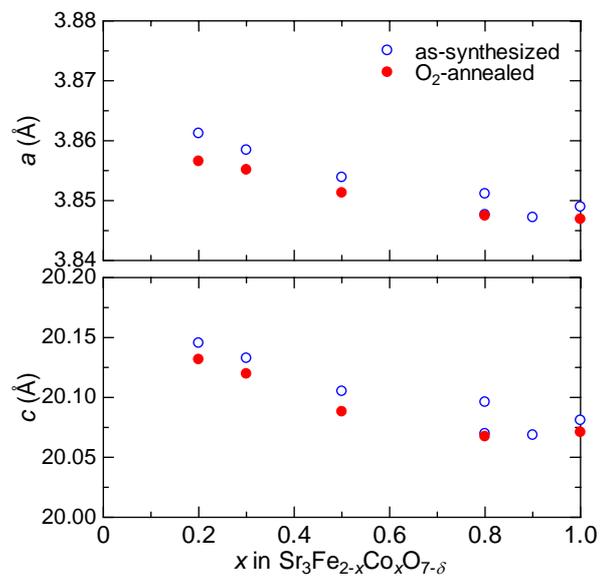

Fig. 2. Motohashi *et al.*

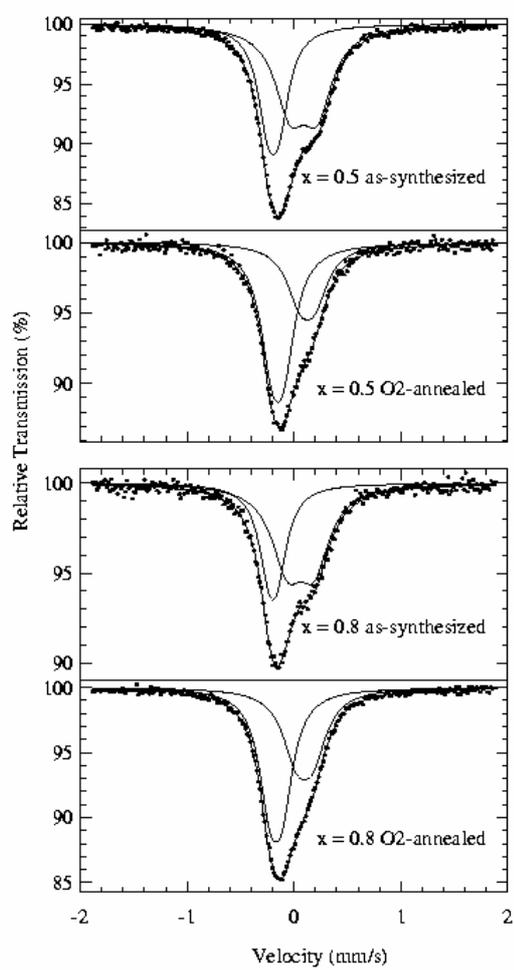

Fig. 3. Motohashi *et al.*

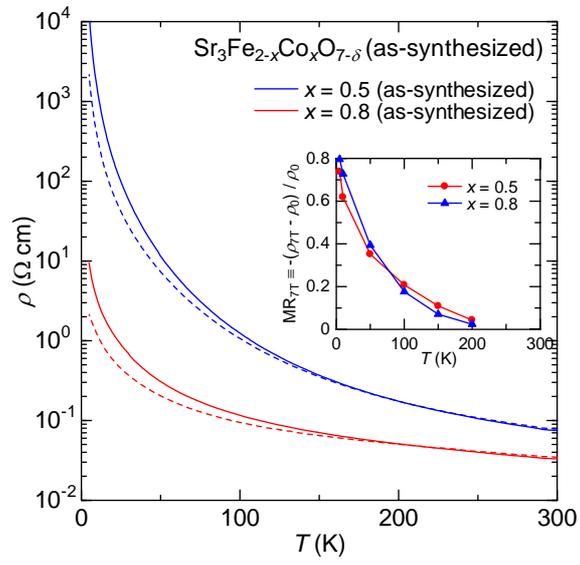

Fig. 4. Motohashi *et al.*

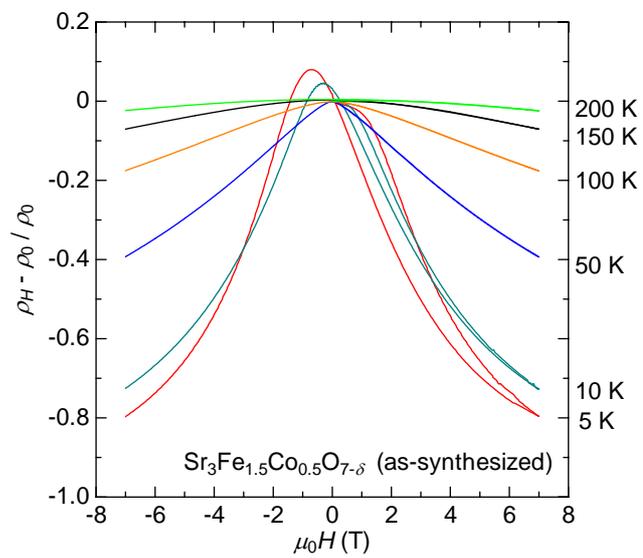

Fig. 5 Motohashi *et al.*

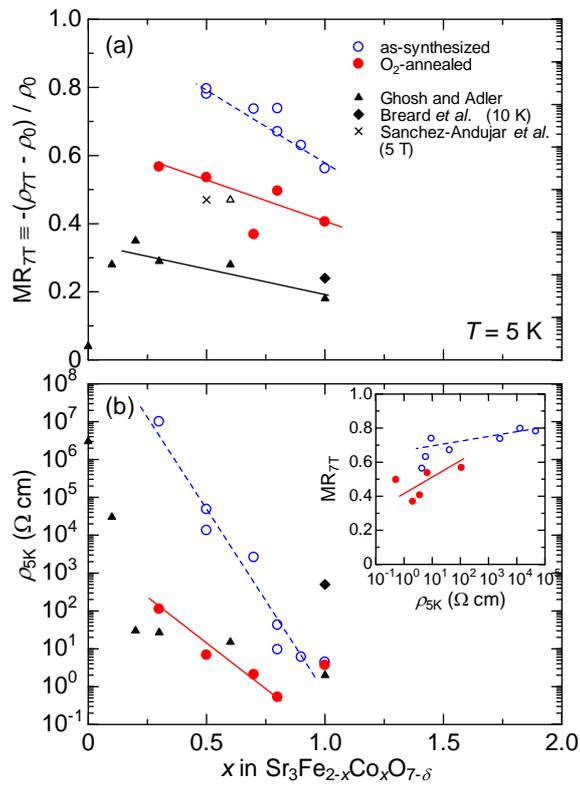

Fig. 6 Motohashi *et al.*

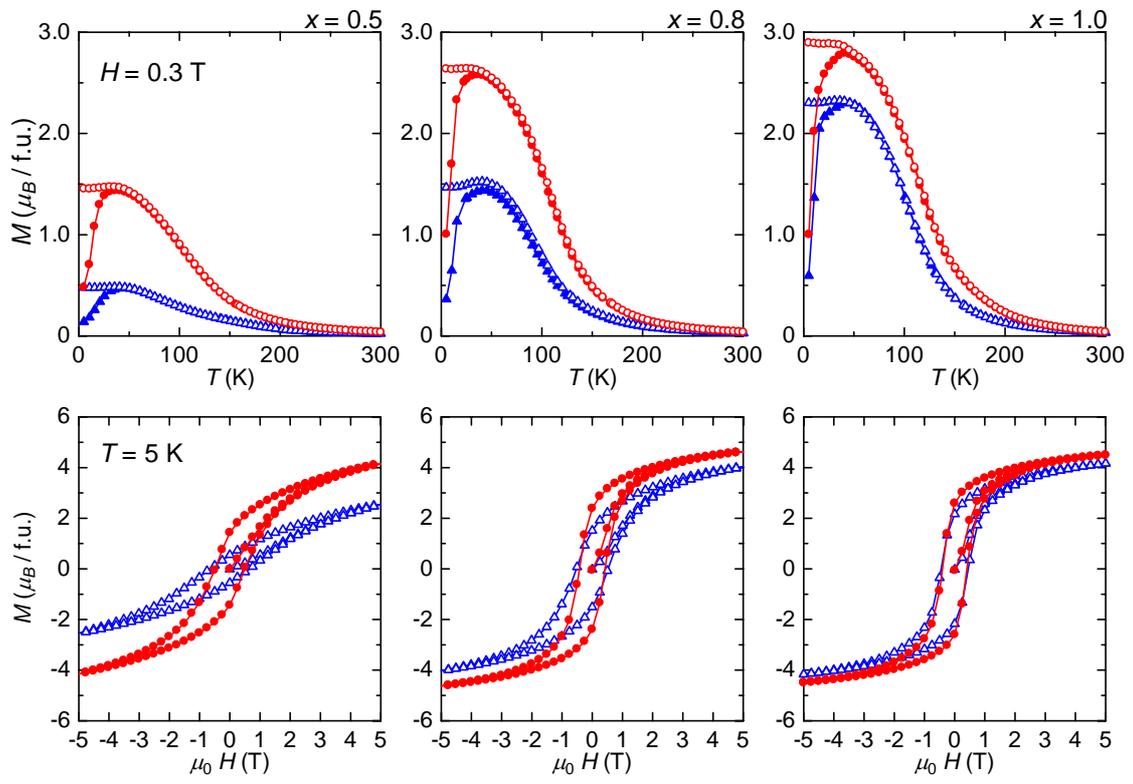

Fig. 7 Motohashi *et al.*

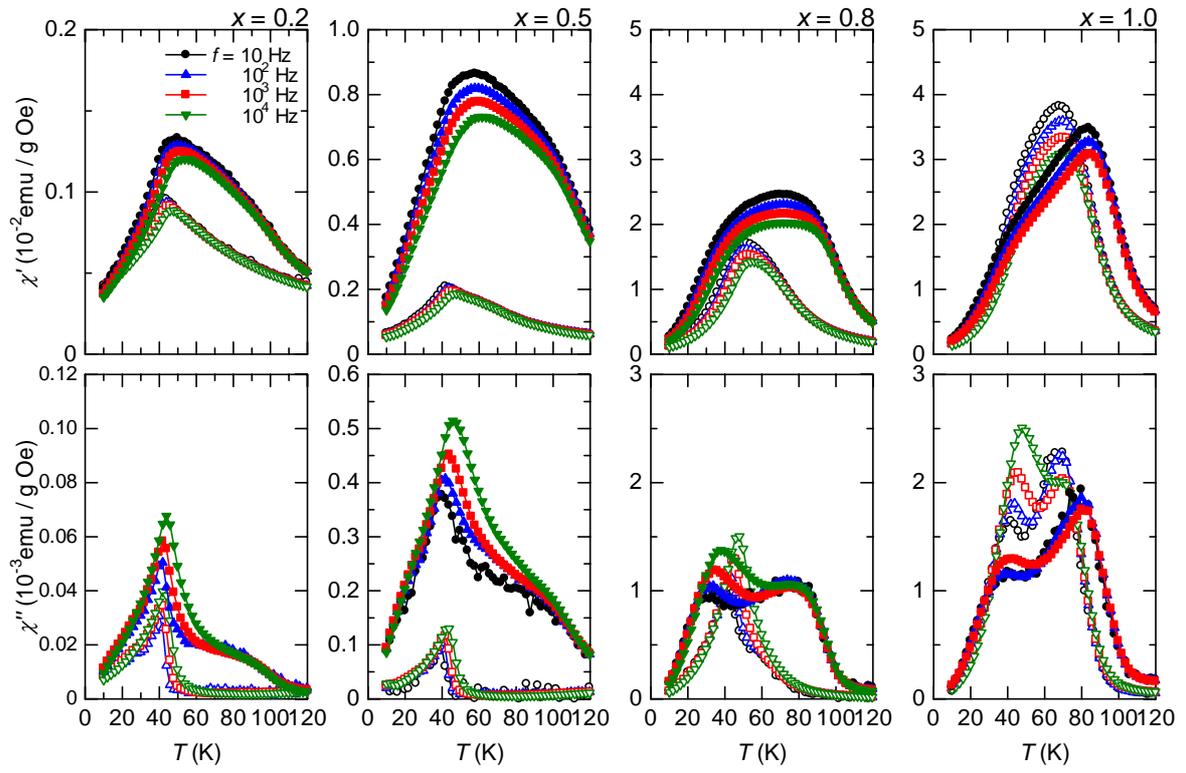

Fig. 8 Motohashi *et al.*

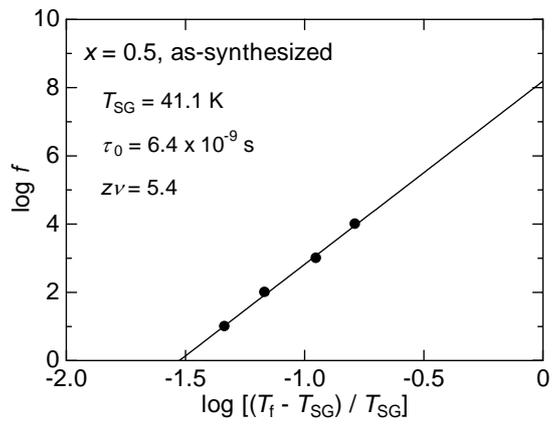

Fig. 9 Motohashi *et al.*

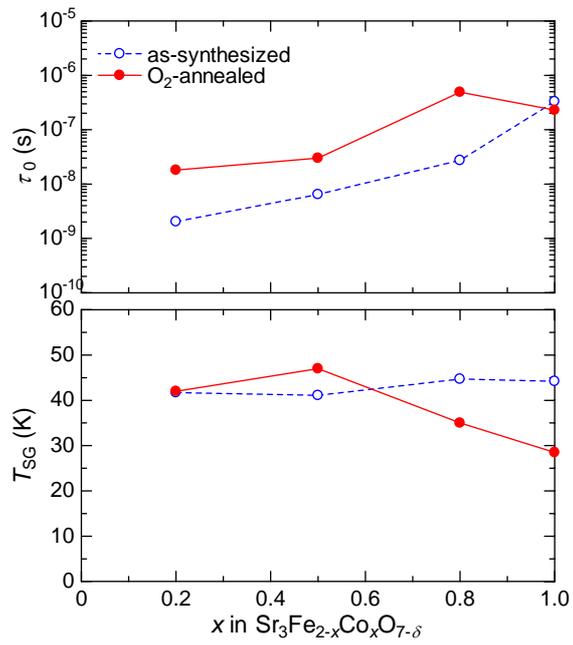

Fig. 10 Motohashi *et al.*